\documentclass[twocolumn,showpacs,preprintnumbers,amsmath,amssymb]{revtex4}
\usepackage{graphicx}% Include figure files
\usepackage{subfigure}
\usepackage{dcolumn}% Align table columns on decimal point
\usepackage{bm}% bold math

\begin{document}

\preprint{APS/123-QED}

\title{Heat Transport in a Random Packing of  Hard Spheres}

\author{Shigenori Matsumoto}
\email{matsumoto@serow.t.u-tokyo.ac.jp}
\affiliation{Department of Applied Physics, Graduate School of Engineering, The University of Tokyo, Bunkyo-ku, Hongo, Tokyo 113-8656.}
\author{Tomoaki Nogawa}
\affiliation{Department of Applied Physics, Graduate School of Engineering, The University of Tokyo, Bunkyo-ku, Hongo, Tokyo 113-8656.}
\author{Takashi Shimada}
\affiliation{Department of Applied Physics, Graduate School of Engineering, The University of Tokyo, Bunkyo-ku, Hongo, Tokyo 113-8656.}
\author{Nobuyasu Ito}
\affiliation{Department of Applied Physics, Graduate School of Engineering, The University of Tokyo, Bunkyo-ku, Hongo, Tokyo 113-8656.}

\date{\today}

\begin{abstract}
Heat conduction in a random packing of hard spheres is studied by nonequilibrium molecular dynamics simulation.
We find a hard-sphere random packing shows higher thermal conductivity than a crystalline  packing with same packing fraction.
Under the same pressure, the random structure causes reduction of thermal conductivity by only 10\% from crystalline packing, 
which is consistent with the experimental fact that amorphous materials can have high thermal conductivity which is comparable to that of crystals.
\end{abstract}
\pacs{Valid PACS appear here}% PACS, the Physics and Astronomy
                             % Classification Scheme.
%\keywords{Suggested keywords}%Use showkeys class option if keyword
                              %display desired
\maketitle

\section{INTRODUCTION}

It is interesting problem how the material structure effects on the transport property.
Amorphous solid often shows lower conductivity than crystals.
For example, the thermal conductivity of silica, which is a glass material used for window glass, is one-eighth of that of quartz, which is the crystal made from the same components.
In other cases, amorphous solids of aluminum nitride show that the maximum thermal conductivity is 85\% of that of pure crystal\cite{aln}.
Such fact indicates the thermal conductivity of amorphous solids can be comparable with that of crystalline solids. 
Similar feature is also found for electric transport in amorphous oxide semiconductors\cite{aos}.

The thermal conductivities of a low-density fluid and a crystal have been studied well rather than amorphous solid. 
In theoretical approaches, the linear response theory well describes the property of macroscopic thermal transport in crystal.
Numerical approaches have been also done to study thermal conductivities of 
fluid and crystal from microscopic dynamics, employing molecular dynamics technique\cite{ogushi,murakami}.
On the other hand, complexities in amorphous structures increase
difficulties of studying thermal properties. 
As one of theoretical approaches, mode-coupling theory has been developed
to treat a localized property which comes from microscopic structures\cite{mct}.
However, the treatment of structural effects is insufficient to fully understand the properties.
Then, numerical approaches are particularly useful to understand macroscopic properties from microscopic structures.
In particular, thermal properties of amorphous solids in nonequilibrium state should be revealed for applications to novel materials. 

Before entering conduction problem, there is a long history of static amorphous structure.
Since 1960s\cite{scott1,scott2,bernal}, structural features of jammed packings have been extensively studied, 
in particular, those of the most dense packing called random close packing (RCP) have been revealed. 
Recently, heat transport in jammed particle packings was studied to
understand the dynamical property of glass\cite{xu,mourzenko, coelho}.
However, mechanisms remain unclear on contributions of such microscopic structures 
to thermal conduction in jammed state. 

\begin{figure}[rt]
  \begin{center}
      \includegraphics[width=9cm]{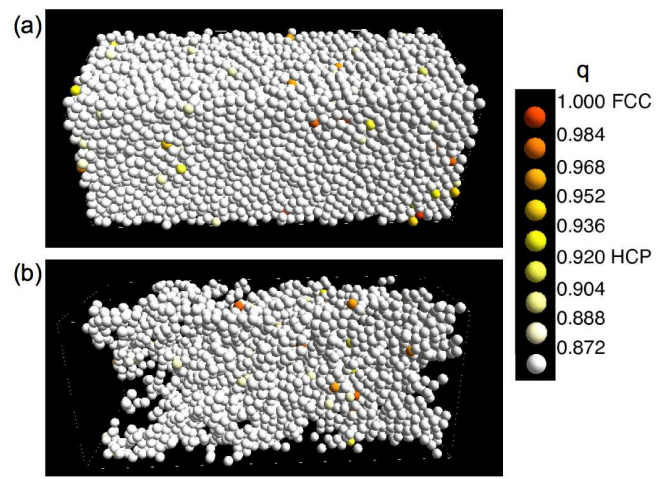}
      \caption{(a) A typical packing structure obtained from our packing method. Colors of the particles correspond to the local order parameter $q$ explained in the body, which shows that the structure is random. (b) The percolating cluster of closely-located particles.}

      \label{randompacking}
  \end{center}
\end{figure}

The purpose of this article is to investigate thermal conductivities of 
jammed packings by the nonequilibrium molecular dynamics simulations.
To examine structural effects on transport, we consider hard-sphere systems.
The hard spheres feel perfect exclusion volume effects of other particles,  which requires special procedures to realize random packings.
Since previously developed method\cite{rcp_JT,rcp_CW,rcp_LS} takes not a small computational times,
 we develop two efficient methods to obtain randomly packed structures.
Further, by imposing temperature gradient, 
we will show that thermal conductivities of random packings
are comparable with those of crystals under same pressures.
We also address that this is due to a characteristic transport paths which enhance high energy-transport in the random structures.

This paper is organized as follows. In the next section, 
we improve the previous packing methods to obtain random packings efficiently.
Then we analyze local structures of the obtained packing in \S \ref{structure}.
Dynamic properties of the packings under thermal gradient  are shown in \S \ref{ht_result}.
The last section is devoted to summary and discussions.

\section{RANDOM PACKING\label{methods}}
\subsection{Model and Methods}

In this section, we show two efficient methods to produce random-packing. 
We consider hard-core particles. The interparticle potential is described by% eq.(1)
\begin{align}
\phi (\boldsymbol r) = \left\{
\begin{matrix}
\infty &r \leq \sigma \\
0 &r>\sigma 
\end{matrix}
\right.,
\end{align}
where $r$ and $\sigma$ denote a distance between particle pairs 
and the diameter of particles, respectively.
Since random-packed structure is not a thermal equilibrium structure,
some artificial procedures are needed to realize it.
It is not a simple task since random putting of particles causes overlapping which is prohibited by hard-core potential.
To solve this some efficient packing methods have been developed\cite{rcp_JT,rcp_CW,rcp_LS}.
Some of the methods take the following procedures. 
We prepare random initial configuration of hard-core particles with sufficiently small diameter.
Then, by increasing the diameter, we obtain dense-packed structures.
In the original procedure, the diameters increase with a constant expansion rate as the simulation time increases.
Therefore, the particles have spaces to expand more in each simulation step.
To expand more efficiently, we adopt two techniques explained below. 
In our simulation, we employ the event-driven molecular dynamics method\cite{isobe}.
In this method, simulation steps proceed by collision events of particles, not by a constant time-step integration.

\subsubsection{Monodisperse packing method}

For the first method, we adopt a variable expansion rate to improve the original method\cite{rcp_LS}.
In this method, the diameters of particles increase uniformly.
We call it ``Monodisperse packing method''.

We consider $L_x \times L_y \times L_z$ 3-dimensional box with periodic boundaries.
In this box $N$ spheres are put randomly. And initial radii are set to be unity.
The initial velocities of particles are randomly assigned from the Maxwell distribution.
We adjust the total velocity of particles to be zero.
The initial state is produced by adding each particle one by one into the simulation box.
If a newly entering particle overlaps with already existing particles, the trial is rejected.
Rejection probability is about 5\%.
Choosing the initial packing fraction to $\phi/\phi_{\rm SC} \sim 0.5$, where $\phi_{\rm SC}$ denotes the volume fraction of the simple cubic packing ($\phi_{\rm SC} = \pi/6 \sim 0.52$). 
We want to expand the diameters of the particles keeping their random initial positions.
To obtain dense packing, however, reconfiguration is necessary because of the hard-core potential.
This is performed by event-driven dynamics.
In accordance with this scheme, we take the following steps: 
\begin{enumerate}
\item Find the minimum gap between the particles $l$ over all the particle pairs.
\item Increase the diameters of the particles uniformly by $xl$, where we introduce a parameter $x (<1)$ as packing speed.
\item Proceed the time until the earliest collision occurs.
\item Translate all particles by half of the next collision time.
\item Repeat the steps 1 to 4 until the packing fraction reaches an aimed value.
\end{enumerate}

\subsubsection{Polydisperse packing method}

Although the above method enables us to produce dense-packing, it takes a lot of calculation cost.
To solve this, we improve the packing process of the former method.
In this second method, the expanding rates are different particle by particle, so we call this ``Polydisperse packing method''.

We first determine a final diameter $\sigma_{\rm final}$.
The initial configuration is obtained by the same way with the first method, and then we take following steps:

\begin{enumerate}
\item Find the local minimum-gap $l_i$ between the particle $i$ ($i=1,..., N$) and surrounding particles.
\item Increase a diameter of particle $\sigma_i$ by $xl_i$. Here, if the new diameter exceeds $\sigma_{\rm final}$, it is set to $\sigma_{\rm final}$. 
\item Perform the same expansion process (steps 1 and 2) over all particles.
\item Repeat the steps 1 and 2 until the new diameters of all particles are calculated.
\item Proceed the time until $N$ collisions occur.
\item Repeat the steps 1 to 4 until diameters of all particles reach $\sigma_{\rm final}$. 
\end{enumerate}

We have introduced the packing speed $x$ in the above two methods.
Choosing $x=0.99$ which is the maximum speed in our simulation, we obtain a random packing shown in Fig.~\ref{randompacking}.
This parameter controls the packing processes from quench to anneal explained below.

\subsection{Packing process}

\begin{figure}[t]
   \begin{center}
       \includegraphics[width=8.5cm]{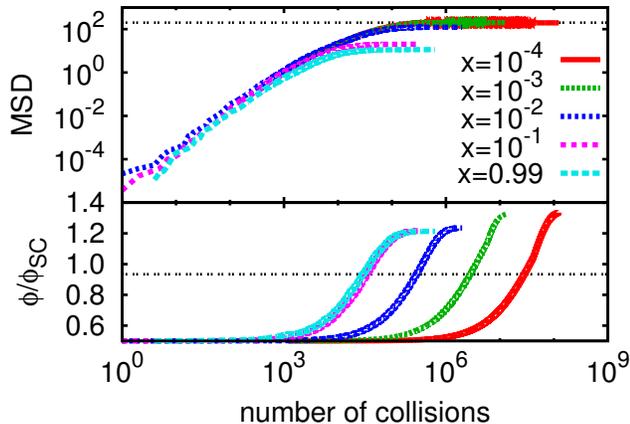}
       \caption{The evolution of the packing fractions (bottom) and MSDs (top) as a function of the number of collisions in the monodisperse packing method. The lines correspond to $\phi_{\rm freeze}$ and the maximum of MSD in the top and the bottom figures.}
       \label{mono_quench_anneal}
   \end{center}
 \end{figure}

We can obtain a homogeneous packing at any simulation time by using the monodisperse packing method.
If the aimed packing fraction is not decided before the simulation starts, the packing finally becomes the closest packing.
On the other hand, when the particles jam, we spend a lot of calculation costs until the jammed packing crystallizes.
To observe the dynamical properties of the monodisperse packing method, we finish each simulation when the maximum of displacement in each cycle becomes less than $10^{-6}$.
Figures \ref{mono_quench_anneal} show the evolution of packing fraction $\phi/\phi_{\rm SC}$ and mean-square displacements (MSD) as a function of the number of collisions per particle by changing $x$ from $0.99$ to $0.0001$ in the monodisperse packing method, respectively.
Here, MSD is defined as an average of square of the difference between the initial and the current positions.
In each simulation the system size is set to $L_x=L_y=20,L_z=40$, and $N=1000$. 

\begin{figure}[t]
  \begin{center}
    \includegraphics[width=8.5cm]{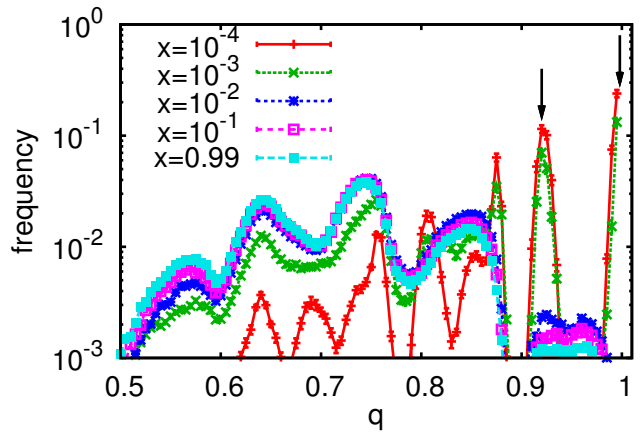}
    \caption{The frequency distribution of the order parameter $q$. The arrows at $q=0.92$ and 1 correspond to the existence of the hexagonal close-packed and the face-centered cubic structures, respectively. }
    \label{q_distribution}
  \end{center}
\end{figure}

For the fastest monodisperse packing $x=0.99$, the displacement of each particle is between twice and triple of the final radius ($\sim 1.34$). 
This means that each particle keeps almost its initial random position.
In this case, the packing fraction approaches $\phi_{\rm RCP}/\phi_{\rm SC} = 1.22$.
This value denotes the packing fraction of randomly closed packing (RCP).
Thus larger values of $x$ correspond to the process of quench.
The displacement continues to grow beyond $2\sigma$ as $x<0.001$, which indicates structural relaxation toward crystallization.
We note that the saturation of MSD at $(L_x^2 + L_y^2 + L_z^2)/12$ is the maximum in the present geometry, thus it does not mean freezing.
When the expanding speeds are sufficiently slow, packing fraction grows beyond $\phi_{\rm RCP}$ and approaches the packing fraction of face-centered cubic (FCC).
This indicates the system  crystallizes.
In this situation, packing structure becomes that of the closest packing, such as FCC and hexagonal closest packing (HCP), which are thermodynamically stable.
In order to distinguish whether the packing is crystal or not, we observe a simple order parameter for $i$-th particle described as
\begin{align}
 q_i = \frac{1}{12}\sum_{j=1}^{n^i_b} |\cos\Theta_j|,\label{order}
\end{align}
where $n^i_b$ denotes the number of  particle in the range of the local gap $l_i<0.15\sigma_{\rm final}$.
This parameter takes 1 for FCC and 0.92 for HCP.
The angle $\Theta_j$ is defined as the maximum bond-angle between $j$-th and other neighboring bonds.
Figure \ref{q_distribution} shows a frequency distribution of $q$.
For $x<0.001$, the peaks at $q=1$ and 0.92 prominently appear.
In this case the packing process corresponds to anneal.
On the other hand, the dominant peaks of the distribution appear below  $q=0.85$ as $x$ increases.
In this case, the packing process corresponds to quench since a local crystalline structure does not appear (see Fig.~\ref{randompacking}(a)).
This crossover corresponds whether the MSD is larger or smaller than $(2\sigma)^2$ when the packing fraction pass through the freezing point, $\phi_{\rm freeze}/\phi_{\rm SC}=0.934$.

 \begin{figure}[t]
   \begin{center}
       \includegraphics[width=8.5cm]{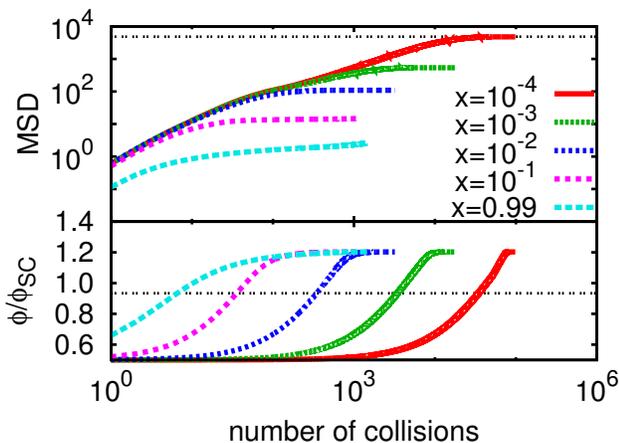}
       \caption{The evolution of packing fractions (bottom) and MSDs (top) in the polydisperse packing method.  The lines correspond to $\phi_{\rm freeze}$ and the maximum of MSD in the top and the bottom figures.}
       \label{poly_quench_anneal}
   \end{center}
 \end{figure}

Similar crossover between anneal and quench is also observed in the polydisperse packing method.
By using this method, the calculation cost is reduced drastically, thus we can obtain larger packings.
Figures \ref{poly_quench_anneal} also show the evolution of $\phi/\phi_{\rm SC}$ and MSD in the polydisperse packing method, respectively.
In each simulation the system size is set to $L_x=L_y=20,L_z=400$, and $N=10000$, which is ten times longer than that used in former method on the $z$-direction.
And we fix the final packing fraction $\phi_{\rm final}/\phi_{\rm SC} = 1.203$ since the packing fraction cannot reach to $\phi_{\rm RCP}/\phi_{\rm SC}$ due to the following reason.
We note here that highly dense packing cannot be obtained by the polydisperse packing method.
In our simulation, it is hard to expand all particles to final diameter for $\phi_{\rm final}/\phi_{\rm SC} > 1.214$ with $x=0.99$.
This is because some particles rapidly expand and freeze in the method, 
and therefore the rest of particles cannot successfully expand their diameters due to the lack of the free space.
While the polydisperse method has huge advantage in the calculation cost,  the method is restricted up to the packing fraction $\phi_{\rm final}/\phi_{\rm SC} = 1.214$ for the fastest packing speed $x=0.99$.

\section{ANALYSIS OF PACKING STRUCTURE\label{structure}}

The packings obtained by the monodisperse and the polydisperse packing method show the same properties analyzed below.
Therefore, the figures in this section will show results only for the polydisperse case.

\begin{figure}[t]
   \begin{center}
       \includegraphics[width=8.5cm]{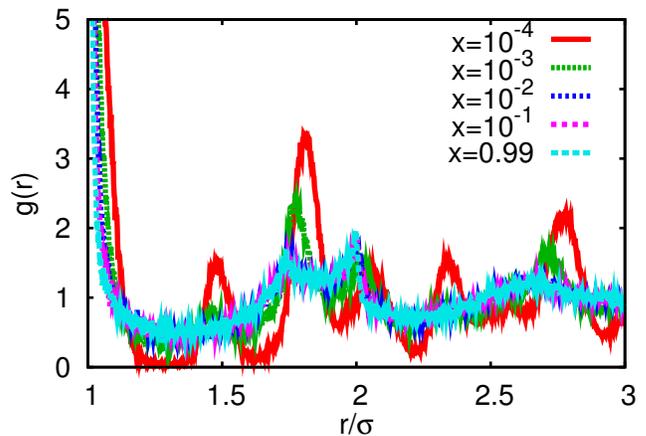}
       \caption{Radial distribution functions for the various packing speeds.}
       \label{rdf_nogawa}
   \end{center}
 \end{figure}

A major signature of random packing structures is the absence of long-range order.
To distinguish the packing structures,  the radial distribution function (RDF) $g(r)$ is calculated.
The function is shown for several values of $x$ in Fig.~\ref{rdf_nogawa}. 
The packings for smaller $x$ show crystalline peaks, those are expected in FCC and HCP structures.
On the other hand, the packings for larger $x$ show only two characteristic 
peaks at $r/\sigma = \sqrt{3}$ and $2$, which is a general feature of RCP\cite{silbert,donev}.
We note that the peak at $r/\sigma=1$ corresponds to particles in contact.

\begin{figure}[t]
   \begin{center}
       \includegraphics[width=8.5cm]{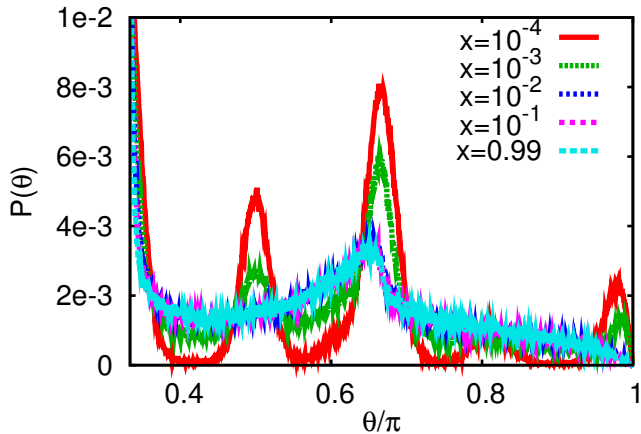}
       \caption{Angular distribution functions for the various packing speeds.}
       \label{adf_nogawa}
   \end{center}
 \end{figure}

\begin{figure}[t]
  \begin{center}
    \includegraphics[width=6cm]{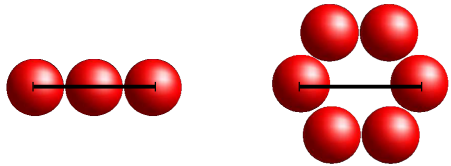}
    \caption{ The rectilinear arrangement and the honeycomb structure. Both structures contribute to the peak at $r/\sigma=2$ in $g(r)$.}
    \label{line_cage}
  \end{center}
\end{figure}

\begin{figure}[t]
    \begin{center}
        \begin{tabular}{c}
            \includegraphics[width=4cm]{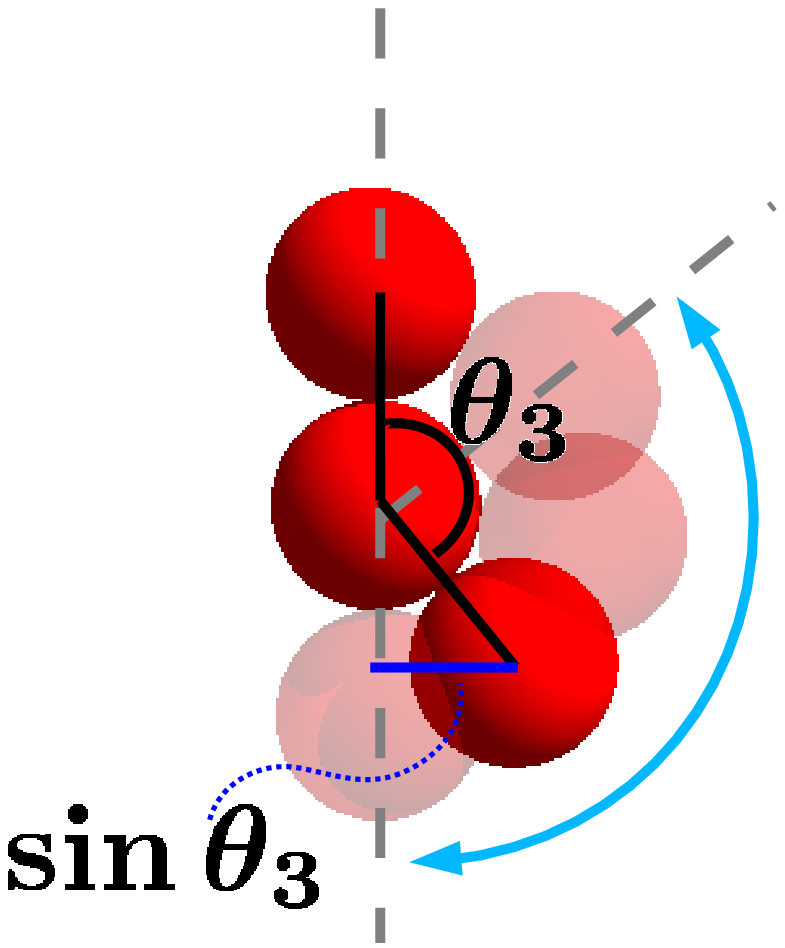}\\
            \includegraphics[width=8.4cm]{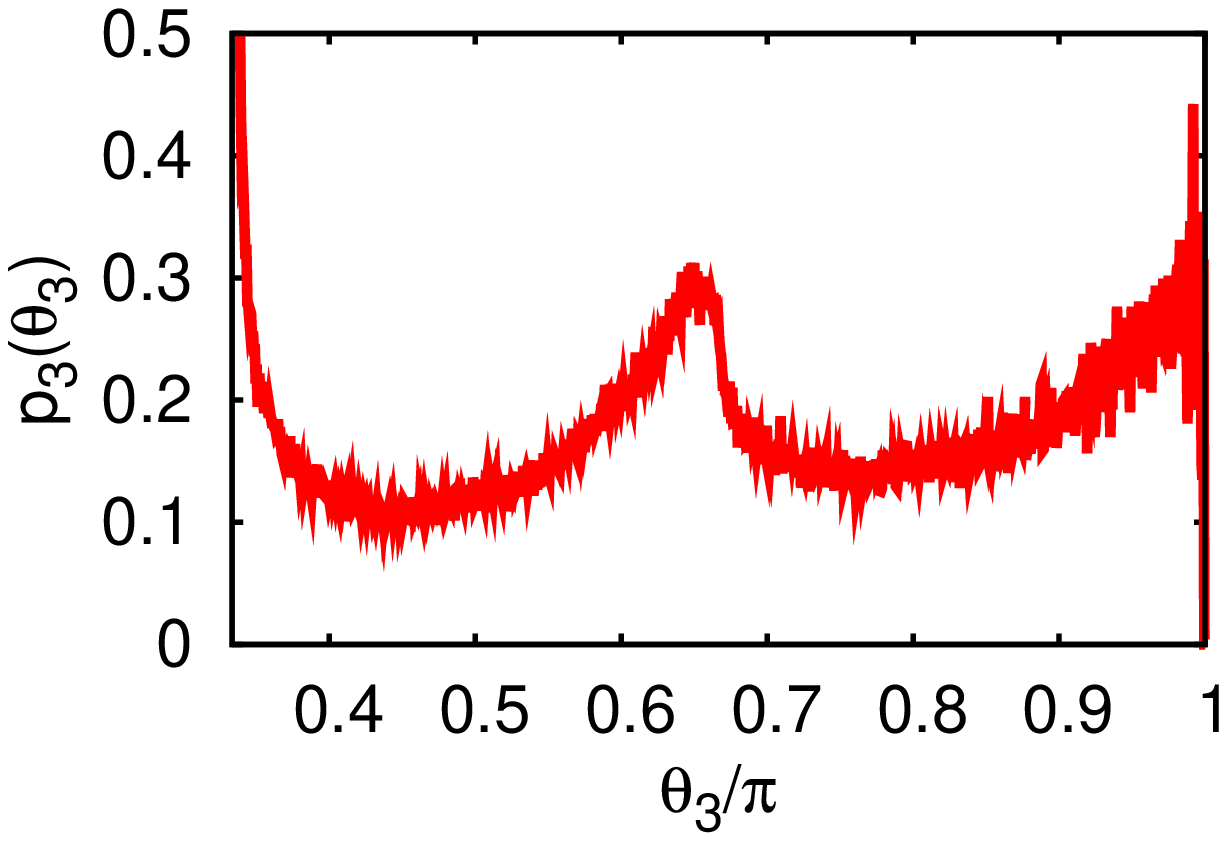} 
        \end{tabular}
    \end{center}
    \caption{ Possible arrangement of three particles in contact (top). Arrow illustrates the ranges of the bond-angle. The reweighted angular distribution function of $\theta_3$ (bottom).}
    \label{reweight3}
\end{figure}

\begin{figure}[t]
    \begin{center}
        \begin{tabular}{c}
            \includegraphics[width=4cm]{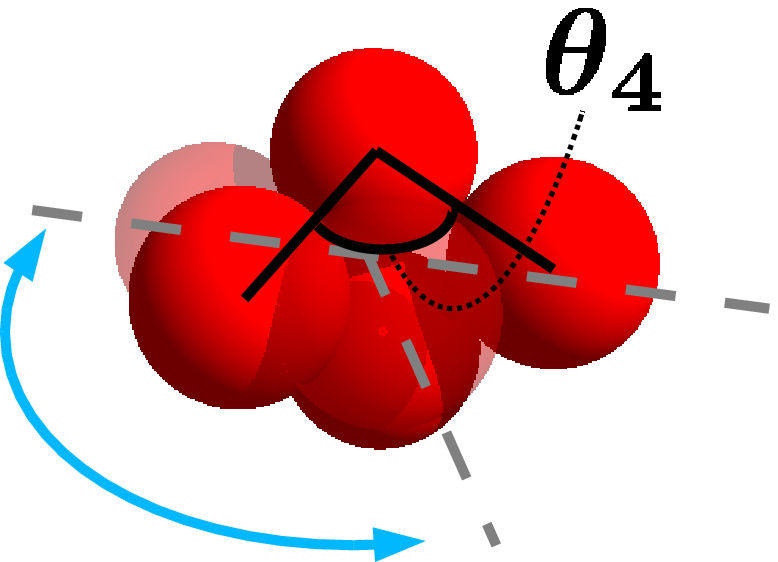} \\
            \includegraphics[width=8.4cm]{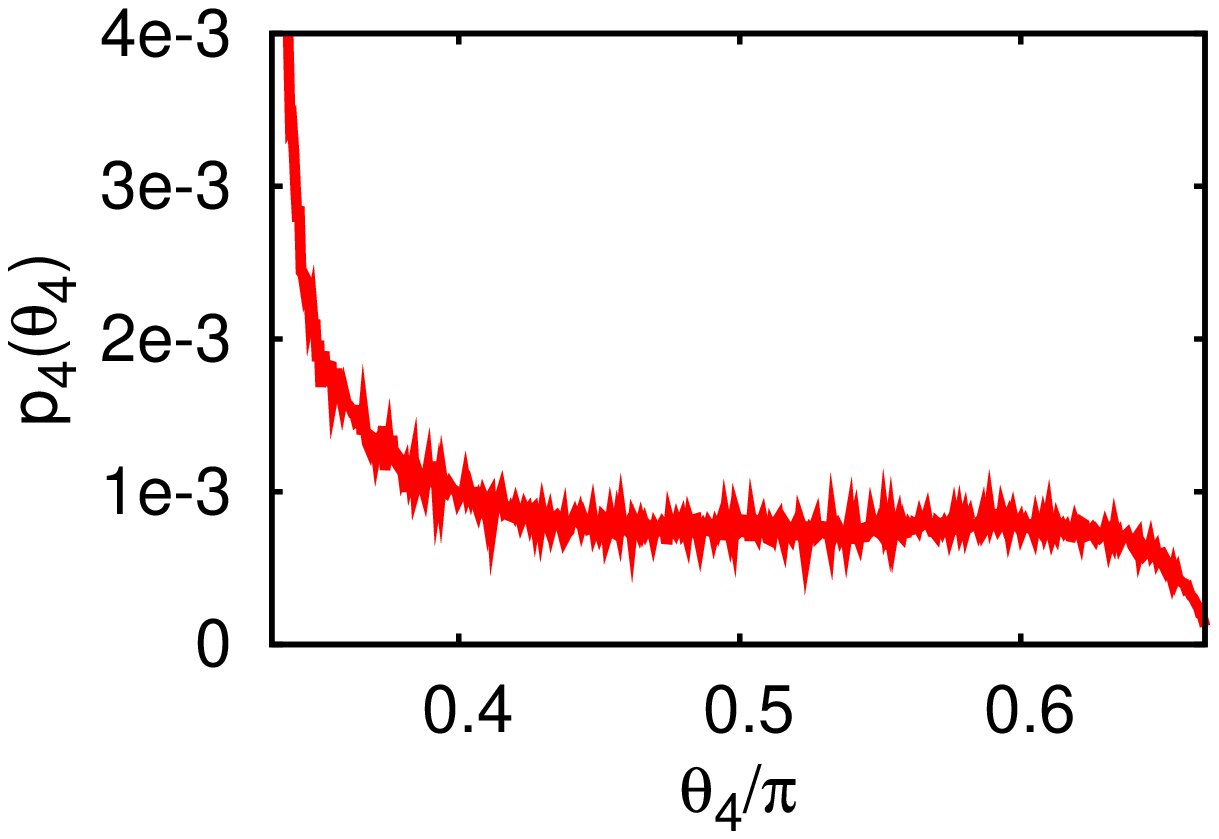} 
        \end{tabular}
    \end{center}
    \caption{ Possible arrangement of four particles in contact (top). Arrow illustrates the ranges of the bond-angle. The reweighted angular distribution function of $\theta_4$ (bottom).}
    \label{reweight4}
\end{figure}

We also calculate angular distribution function $P(\theta)$ shown in Fig.~\ref{adf_nogawa},
where $\theta$ denotes the bond-pair angle.
We find following two characteristic features on $g(r)$ and $P(\theta)$ of quenched packings:
While the peak at $r/\sigma = 2$ appears in $g(r)$, 
rectilinear arrangement (see Fig.\ref{line_cage} (left)) seems not to exist from $P(\theta=\pi)\simeq 0$.
This fact seems to imply this peak comes from another structure, e.g. cage structures such as the honeycomb structures (see Fig.~\ref{line_cage} (right)).
While the other characteristic peaks of crystals do not appear, 
the peak at $r/\sigma = \sqrt{3}$ remains.
These features have been also observed in previous study\cite{donev}, but those reasons have not been mentioned.
To clarify the origins of those peaks, we consider two cases of contacting particles 
and their spatial contacting bond-angles illustrated in Fig.~\ref{reweight3} and \ref{reweight4}. 

As in Fig.~\ref{reweight3} (top), we consider three particles.
Possible positions for the third particle around
the two contacting particles are restricted in the range $\pi/3 < \theta_3 < \pi$.
If particles distribute homogeneously in that range, the probability of finding $\theta_3$, $f_3(\theta_3)$, is proportional to $\sin\theta_3$ illustrated  in Fig.~\ref{reweight3} (top).
In this three-particles case, we obtain a reweighted distribution $p_3(\theta_3)$ described as
\begin{align}
p_3(\theta_3) &\equiv P(\theta_3)f_3(\theta_3)^{-1} \qquad \frac{\pi}{3} < \theta_3 < \pi\nonumber\\
f_3(\theta_3) &= 2\pi \sin(\theta_3).
\end{align}
This reweighting also tells us why we see no peak at $\theta=\pi$ (see Fig.~\ref{reweight3} (bottom)).

Similarly, we consider four-particles unit as in Fig.~\ref{reweight4} (top).
We assume triangular arrangements of three particles and then consider positions of fourth particle.
Since the triangular arrangements corresponds to $\theta_3 = \pi/3$ in 
the three-particles case, the assumption is reasonable 
that such arrangement will be frequently observed in the packing.
We obtain the reweighted distribution $p_4(\theta_4)$ as the following expression:
\begin{align}
p_4(\theta_4) &\equiv P(\theta_4)f_4(\theta_4)^{-1} \qquad \frac{\pi}{3} < \theta_4 < \frac{2\pi}{3}\nonumber\\
 f_4(\theta_4) &= \frac{\sin\theta_4}{\sqrt{(1-\cos\theta_4)(1/2+\cos\theta_4)}}.
\end{align}
As shown in Fig.~\ref{reweight4} (bottom), the reweighted distribution $p_4(\theta_4)$ 
is approximately flat in the region $\pi/3 < \theta_4 < 2\pi/3$ .
This result indicates that the peak structure we see in $P(\theta)$ is solely the result of the reweighting probability in the region $\pi/3 < \theta_4 < 2\pi/3$ .

We have found the characteristic local structure of random packing.
In our analysis, the random packings obtained by two our methods show the same random structure.
Therefore,  in the following section, we show results 
obtained only by the polydisperse packing method 
because of its computational inexpensiveness.

\section{HEAT TRANSPORT IN RANDOM PACKING\label{ht_result}}

\subsection{Simulation settings}
We investigate thermal properties of the random packing and compare to those of crystals.
As the crystalline structure, 
we adopt FCC, which is the most stable structure in three-dimensional hard-core system.
Here, we note that the thermal conductivity highly depends on the packing fraction.
Therefore, we use a little loose packings both for FCC and random ones and compare them
by setting the density to same.

We impose periodic boundary conditions in $x$- and $y$- directions.
And we set heat walls on the both sides of $z$-direction in the following way.
At first, we make homogeneous packings with periodic boundary condition.
And then particles in the region $z=0$ to $2\sigma/\sqrt{3}$ and $z=L_z-2\sigma/\sqrt{3}$ to $L_z$ are regarded as a part of heat walls 
and those are fixed throughout the simulations.
We choose (110) surface for FCC as heat walls.
When a free particle collides with particles in each wall, 
the particle bounces back with a new velocity randomly chosen from thermal equilibrium distribution of each temperature.
The velocity distribution with temperature $T_{\rm B}$ is described as
\begin{align}
f(v_{\rm n}, v_{\rm t,1}, v_{\rm t,2}) &= \phi(v_{\rm n})\psi(v_{\rm t,1})\psi(v_{\rm t,2})\\
\psi(v) &= \frac{1}{\sqrt{2\pi k_{\rm B}T_{\rm B}}} \exp\left( -\frac{v^2}{2k_{\rm B}T_{\rm B}}\right)\\
\phi(v) &= \frac{1}{k_{\rm B}T_{\rm B}}|v| \exp\left( -\frac{v^2}{2k_{\rm B}T_{\rm B}}\right),
\end{align}
where $v_{\rm n}, v_{\rm t,i}$ denote a normal vector and orthonormal tangential-vectors on a colliding point of the wall particle.
The Boltzmann constant $k_{\rm B}$ is chosen to be unity. 
By employing similar heat bath, Murakami {\it et al.}\cite{murakami} showed that 
the hard-core fluid systems produce the Fourier-type heat conduction.
They also investigate the system-size dependence of thermal conductivity, 
which is consistent with theoretical predictions of the Kubo formula and 
long-time tail of autocorrelation function of heat flux.

\subsection{Definition of physical quantities}

Our system attains a nonequilibrium steady state,
where heat steadily flows from high-temperature to low-temperature sides;
 $T_{\rm H}$ around $z=0$ and $T_{\rm L}$ ($T_{\rm H} > T_{\rm L}$) around $z=L_z$, respectively.
In the steady state, it is reasonable to assume local equilibrium\cite{murakami}, and  then temperature is defined as
\begin{align}
T(z)& = \frac{1}{d}\left.\left\langle\sum_{i \in B(z)} \boldsymbol v_i^2\right\rangle \right/\left\langle \sum_{i \in B(z) } 1 \right\rangle
\end{align}
where $d$ and $\boldsymbol v$ denote spatial dimension and velocity vector of $i$-th particle, respectively.
The bracket denotes time average.
And $B(z)$ means a group of particles exist in the region, $(z-\delta/2,z + \delta/2)$.

\begin{figure}[!h]
    \includegraphics[width=9cm]{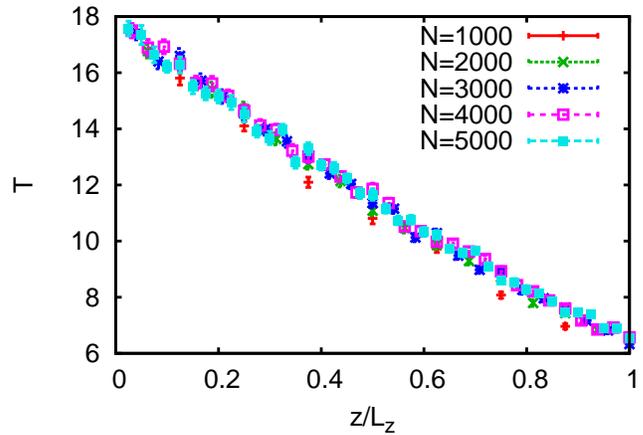}
    \caption{Temperature profiles in the random packing with different system sizes.}
\label{temp_dens}
\end{figure}

As the same as observed in the low-density fluid case\cite{murakami} $\phi/\phi_{\rm SC}=0.69$,
 present random packings also exhibit linear temperature profiles (Fig.~\ref{temp_dens}).
The well-scaled linear profile satisfies the necessary condition to evaluate the thermal conductivity.
Heat conduction generally comes from energy and mass transports.
In the dense packing state, energy transport by collisions mainly contributes to the thermal conductivity.
Since all particles almost cannot move from the initial positions, the contribution from mass transports can be neglected.

In the steady state, we only measure heat flux by the energy received from
the high-temperature heat wall since energy conserves in the bulk of our system.
Thus the heat flux is defined as
\begin{align}
Q(t_n) &= \sum^n_{k=1} (\Delta E)_k,
\end{align}
where $(\Delta E)_k$ denotes the received energy by $k$-th collision with the wall, 
and $t_n$ denotes the total elapsed time after $n$ collisions with the wall.
Thermal conductivity is defined as
\begin{align}
\kappa(L_z) =  - \frac{\langle J_z\rangle}{\nabla T},
\end{align}
where $J_z=Q(t_n)/t_n$.

\subsection{Results}

\begin{figure}[t]
\begin{center}
\includegraphics[width=8.5cm]{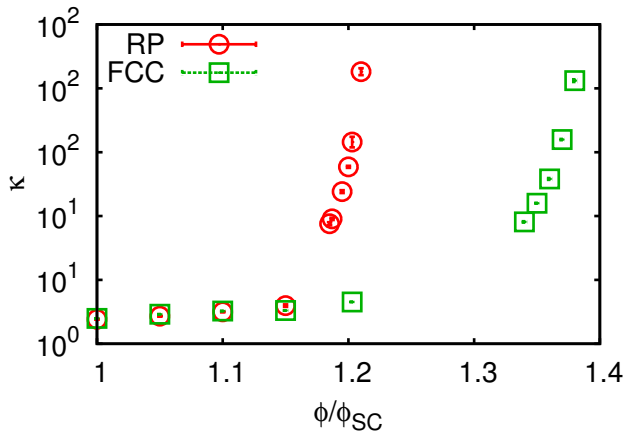}
\caption{Thermal conductivities of the random packing (RP) and FCC with $N=6000$ against the packing fraction. }
\label{dvt}
\end{center}
\end{figure}

Figure \ref{dvt} shows the packing fraction dependence of thermal conductivity both for random packing and FCC.
In this simulation, we fix the temperature ($T_{\rm H}=18, T_{\rm L}=6$) and 
the box size is $L_z = 240, L_x=L_y=20$ for 6000 particles.
Below $\phi/\phi_{\rm SC} = 1.15$, the random packing rapidly crystallizes since the particles have enough space for reconfiguration.
Typically, such crystalline nucleations occur near the high-temperature wall and then it grows.
This phenomena have been also observed experimentally using rigid particles\cite{hard_crystal}.
Actually, crystal growth technique under thermal gradient is widely used in the field of engineering.
This technique  was also applied in colloid system\cite{colloid}. 
In the range $\phi/\phi_{\rm SC} > 1.15$, thermal conductivities of the random packing and FCC shows diverging behavior
around distinct packing fractions of RCP and the closest packing, respectively.
Note that, in random packing simulations, the system crystallizes with a small probability even near the RCP point.

\begin{figure}[t]
\begin{center}
\includegraphics[width=8.5cm]{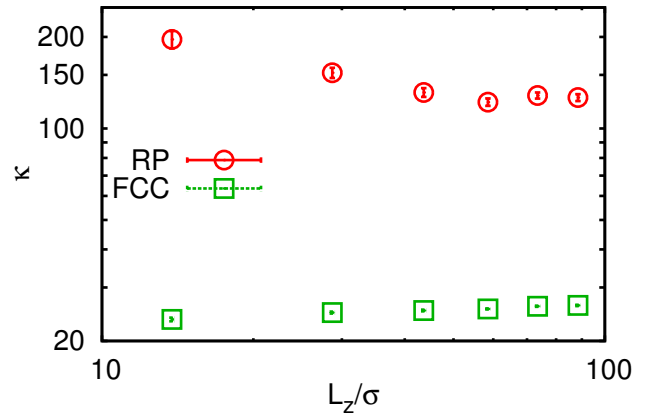}
\caption{System size dependence of the thermal conductivities of the random packing (RP) and FCC.}
\label{thermal_conductivity}
\end{center}
\end{figure}

We also investigate the system size dependence of the thermal conductivity 
at a constant packing fraction $\phi/\phi_{\rm SC} = 1.203$, which is slightly smaller than RCP fraction, as shown in Fig.~\ref{thermal_conductivity}.
We find that thermal conductivity of random packing is far larger than FCC, which is discussed later.
In the FCC $\kappa$ is proportional to $L_z^{-1/2}$ which is consistent 
with linear response theory, i.e. Kubo formula, in which thermal conductivity is described as
\begin{align}
\kappa = \lim_{t\to\infty}\lim_{V\to\infty}\frac{1}{Vk_{\rm B}T^2}\int^t_0 dt'\langle J(0)J(t')\rangle.
\end{align}
And autocorrelation function is supposed to have a slow decay $t^{-d/2}$, 
which is commonly called long-time tails in hard-core particle system\cite{alder2,alder3}.
Thus the thermal conductivity for finite size system shows following 
size dependence;
\begin{align}
\kappa (L_z) \sim 
\left\{
\begin{array}{cc}
\log L_z  & \text{in 2D}\\
L_z^{-1/2} & \text{in 3D} \label{size_dependence}
\end{array}\right..
\end{align}

\begin{figure}[t]
\begin{center}
\includegraphics[width=8.5cm]{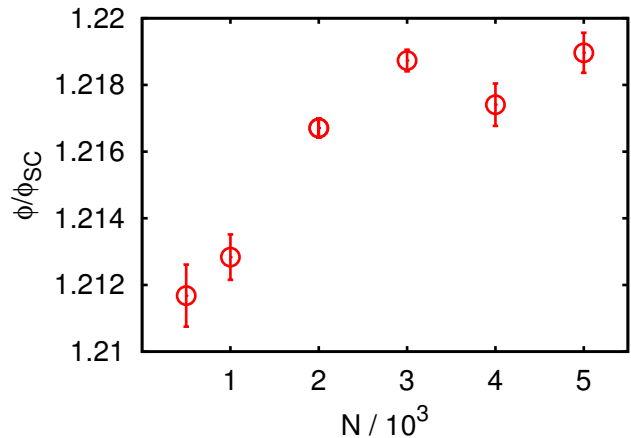}
\caption{Maximum packing fractions obtained from the monodisperse packing method as a function of the system size.}
\label{size_density}
\end{center}
\end{figure}

On the other hand, the thermal conductivity of the random packing 
 decreases as $L_z$ increases (see Fig.~\ref{thermal_conductivity}), which is opposite to the above prediction.
This is explained by the interparticle distance. 
In the smaller systems, it becomes more difficult to obtain dense random packing around RCP.
Figure \ref{size_density} shows obtained packing fractions by the monodisperse packing method explained in section \ref{methods}.
For $N<3000$, the obtained random packings show lower fraction than $\phi/\phi_{\rm SC}\approx 1.218$, which is universally obtained for lager systems.
Since the effective packing fraction of RCP becomes low in these smaller systems, the obtained random packing are effectively closer to RCP. 
Therefore the interparticle distance becomes short. 
On the other hand, the interparticle distance becomes larger as system size increases.
This effective looseness causes lower conductivity than that of smaller systems.

\begin{figure}[!h]
\begin{center}
\includegraphics[width=8.5cm]{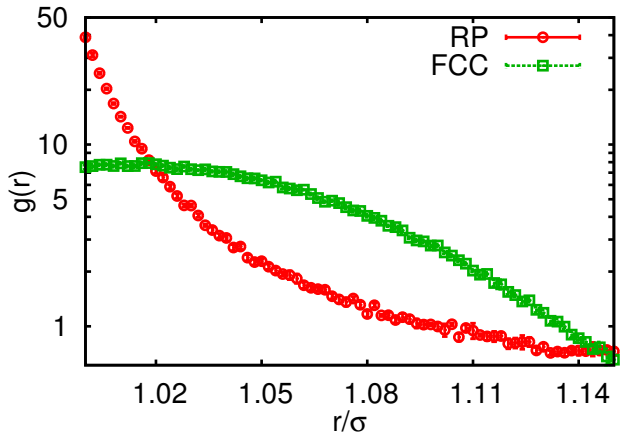}
\caption{Radial distribution functions for the random packing (RP) and FCC around $r/\sigma = 1$ with the same packing fraction.}
\label{rdf_near1}
\end{center}
\end{figure}

High conductivity of random packing can be explained by a large number of particle collisions
per unit time, which enhance energy transport.
Roughly, collision frequency of particle pair is inversely proportional to interparticle distance.
Actually, the distances in random packing are shorter than that of FCC.
Figure \ref{rdf_near1} shows $g(r)$ around $r/\sigma = 1$.
In  FCC $g(r)$ shows a Gaussian distribution around the average of the interparticle distance.
On the other hand, a diverging peak appears at $r/\sigma = 1$ in the random packing.
This fact implies that such close-particle bonds form a efficient path for thermal conduction.
As a consequence of this, a single large cluster percolating from the one side to the other in the $z$-directions is observed by assuming that neighbor particles within the range $r/\sigma <1.006$ belong to same cluster shown in Fig.~\ref{randompacking}(b).
In case of FCC, a such cluster is only located in the low-temperature region and does not bridge the space between two thermal baths.

\begin{figure}[!h]
   \begin{center}
    \includegraphics[width=8cm]{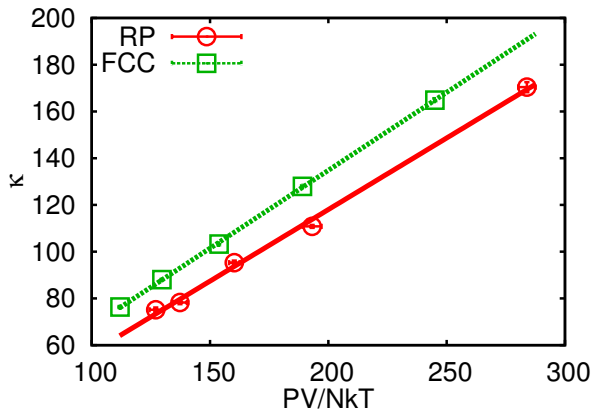}
    \caption{Pressure dependence of the thermal conductivity.}
\label{p_tc}
\end{center}
\end{figure}

We also investigate pressure dependence of thermal conductivities with $N=6000$. 
The size effect of thermal conductivities is negligible in this condition.
In Fig.~\ref{p_tc}, thermal conductivities are plotted against 
pressures calculated in an equilibrium state. 
Compared at same pressure, the thermal conductivity of FCC is higher than that of the random packing in contrast to the case of same density condition.
Thermal conductivity of random packing is, however, reduced only by 10\% of that of FCC.
This difference is considered to come from the nature of the paths of sequential collisions which is relatively straight in FCC and wondering in the random packing.

\section{CONCLUSION}

We investigate heat transport properties of random packings of hard spheres 
by nonequilibrium molecular dynamics simulations.

We introduce two efficient methods to obtain random packings.
In the monodisperse packing method,  we can produce dense random packings homogeneously.
In the polydisperse packing method, random packings can be obtained quickly in the density range $\phi/\phi_{SC} < 1.214$.
Both methods have a parameter $x$ which controls the packing speed.
When this speed is so fast particles cannot diffuse sufficiently before freezing, hence the packing structure becomes random.
The obtained structure is recognized by analyzing the local crystallization parameter and the radial and the angular distribution functions whether the packing is random  or the crystalline.

Using the obtained random packing, we investigate heat transport properties by imposing parallel heat walls at the both ends.
And then we compare the thermal conductivity with that of the crystalline packings.
Compared at the same density, the conductivity of the random packing is higher than that of the crystal 
since the percolated cluster of closely-located particles exists in the system.
On the other hand, compared at the same pressure, the conductivity of the random packing is about 10\% smaller than that of the crystal.
These results suggest that the amorphous solids can have comparable thermal conductivity to crystals.
The mechanism of the enhancement of the thermal conductivity by the percolated cluster is also expected in other kinds of transports, such as electric conduction.

\section{Acknowledgments}
We thank Dr.Atsushi Kamimura for helpful discussions.
This work was partly supported by Award No. KUK-I1-005-04 made by King Abdullah University of Science and Technology (KAUST). 

\newpage %Just because of unusual number of tables stacked at end

\bibliography{randompack}% Produces the bibliography via BibTeX.

\end{document}